\documentclass[letter,traditabstract]{aa} 
\usepackage{graphicx}
\usepackage{txfonts}
\usepackage{deluxetable}
\def\ga{\mathrel{\raise0.35ex\hbox{$\scriptstyle >$}\kern-0.6em
\lower0.40ex\hbox{{$\scriptstyle \sim$}}}}
\def\la{\mathrel{\raise0.35ex\hbox{$\scriptstyle <$}\kern-0.6em
\lower0.40ex\hbox{{$\scriptstyle \sim$}}}}
\def\co{CO {\it J}=1-0 }
\def\cotwo{CO {\it J}=2-1 }

\def\cofive{CO {\it J}=5-4 }
\def\hij{high-{\it J} }

\begin{document}
   \title{[CII] line emission in BRI~1335-0417 at z=4.4\thanks{Based on data collected with the 
  ESO APEX telescope under program ID 084A-0030}}

   \author{Jeff Wagg\inst{1} 
               \and  Chris L. Carilli\inst{2} 
              \and David J. Wilner\inst{3} \and Pierre Cox\inst{4} \and Carlos De Breuck\inst{5}
              \and Karl Menten\inst{6} \and  Dominik A. Riechers\inst{7,9} \and Fabian Walter\inst{8}
          }

   \institute{European Southern Observatory, Casilla 19001, Santiago, Chile;
            jwagg@eso.org
  \and National Radio Astronomy Observatory, P.O. Box 0, Socorro, NM, 87801
  \and Harvard-Smithsonian Center for Astrophysics, 60 Garden Street, Cambridge, MA 02138, USA
  \and Institute de Radioastronomie Millimetrique, St. Martin d'Heres F-38406, France
  \and European Southern Observatory, Karl-Schwarzschild-Str. 2 D-85748, Garchingbei M\"{u}nchen, Germany
  \and Max-Planck-Institut f\"{u}r Radioastronomie, Auf dem H\"ugel 71, 53121 Bonn, Germany
  \and California Institute of Technology, 1200 East California Boulevard, Pasadena, CA 91125, USA
  \and Max-Planck-Institute for Astronomy, K\"{o}nigsstuhl 17, 69117 Heidelberg, Germany
  \and Hubble Fellow
             }

   \date{accepted}

  \abstract{Using the 12m APEX telescope, we have detected redshifted emission from the 157.74$\mu$m [CII] line in the $z=4.4074$ quasar BRI~1335-0417. The linewidth and redshift are in good agreement with previous observations of \hij CO line emission. We measure a [CII] line luminosity, 
 $L_{[CII]} = (16.4 \pm 2.6)\times 10^9$~$L_{\odot}$, making BRI~1335-0417 the most luminous, unlensed [CII] line emitter known at high-redshift. The [CII]-to-FIR luminosity ratio of $(5.3 \pm 0.8) \times$10$^{-4}$ is $\sim$3$\times$ higher than expected for an average object with a FIR luminosity $L_{FIR} = 3.1\times 10^{13}$~$L_{\odot}$, if this ratio were to follow the trend observed in other FIR-bright galaxies that have been detected in [CII] line emission. These new data suggest that the scatter in the [CII]-to-FIR luminosity ratio could be larger than previously expected for high luminosity objects. BR1335-0417 has a similar FIR luminosity and [CII]/CO luminosity compared to local ULIRGS and appears to be a gas-rich merger forming stars at a rate of
a few thousand solar masses per year.}

   \authorrunning{J. Wagg et al.}
   \titlerunning{[CII] in BRI~1335-0417}

   \maketitle
%

\section{Introduction}

The dominant interstellar medium (ISM) gas cooling line in star-forming galaxies is the 157.74~$\mu$m fine structure 
line of [CII]~$^2P_{3/2} \rightarrow ^2P_{1/2}$, which typically exhibits a luminosity $\sim$0.1--1\%
that of the far-infrared (FIR) 
luminosity (rest-frame 42.5 to 122.5~$\mu$m) in quiescent galaxies, and less than 0.1\% in ultra-luminous
 infrared galaxies (ULIRGs; Crawford et al.\ 1985; Stacey et al.\ 1991; 
 Wright et al.\ 1991; Malhotra et al.\ 1997). The bulk of [CII] line emission
 arises in warm ($\ge$200~K) and dense ($n_{cr} = $~3$\times$10$^{3}$~cm$^{-3}$) gas 
associated with photo dissociation regions (PDRs) found at the
 surface between neutral and ionized gas clouds.  
[CII] line emission therefore provides a powerful means of studying
 star-formation and gas kinematics in galaxies. 

Although it is not possible to detect the [CII] emission line in nearby galaxies using ground-based facilities,  
 successful observations of this line at $z \gg 1$ have been achieved with submm/mm-wavelength telescopes.
 The first detection of [CII] line emission in the $z=6.4$  quasar host galaxy, J1148+5251, was made using the IRAM 30m telescope (Maiolino  et al.\ 2005), confirming that the [CII]-to-FIR luminosity ratio is decreasing in such hyper-luminous infrared galaxies (e.g. Malhotra et al.\ 2001; Maiolino et al.\ 2009). Subsequent interferometric imaging with the IRAM Plateau de Bure Interferometer (PdBI) by Walter et al.\ (2009) showed that [CII] emission traces a large burst of star-formation over physical scales of $\sim$1.5~kpc, a size nearly an order of magnitude larger than that of the nuclear starbursts observed in nearby galaxies like
 Arp220.  This emission line has also been detected in other  high-redshift quasar host galaxies, including BR1202-0725 at $z = 4.7$
with the SMA (Iono et al.\ 2006), and in the lensed object BRI~0952-0115 at $z=4.4$ with the Atacama Pathfinder Experiment (APEX) 
telescope (Maiolino et al.\ 2009). Similarly,
 strongly lensed luminous infrared starburst galaxies have been detected in [CII] line emission, including a CSO detection of [CII] in  
MIPS~J1428+3526 at $z=1.3$ (Hailey-Dunsheath et al.\ 2010) and a tentative detection in SMM~J2135-0102 at $z = 2.3$ with  \textit{Herschel} (Ivison et al.\ 2010). 

BRI~1335-0417 is a quasar host galaxy at $z = 4.4$ whose AGN emits weak Ly$\alpha$, N~V and C~IV line emission  (Storrie-Lombardi et al.\ 1996), and has a 
 likely black hole mass of  6$\times$10$^9$~$M_{\odot}$ (Shields et al.\ 2006).
   Thermal dust continuum emission detected at 1.25~mm with the IRAM~30m (Omont et al.\ 1996) suggests the presence of a significant 
quantity of dust ($\sim$10$^{9}$~M$\odot$), which shorter wavelength 350~$\mu$m observations constrain to have a  temperature 
 of $\sim$43~K (Benford et al.\ 1999). The implied far-infrared luminosity of $3.1\times$10$^{13}$~$L_{\odot}$ is consistent with a massive burst of star-formation in a heavily obscured system, one which is being fueled by  $\sim$9$\times$10$^{10}$~$M_{\odot}$ of warm molecular gas,
 as indicated by the presence of \cotwo and \cofive line emission (Guilloteau et al.\ 1997; Carilli et al.\ 1999, 2002; Riechers et al.\ 2008). In spite of its 
tremendous apparent luminosity, BRI~1335-0417 does not appear to be gravitationally lensed (Storrie-Lombardi et al.\ 1996), but is rather a massive 
 galaxy in the process of  formation via a gas-rich, or `wet' merger (as revealed by spatially resolved CO imaging by Riechers et al.\ 2008).

Here, we present APEX\footnote{APEX is a collaboration between the 
Max-Planck-Institut f\"{u}r Radioastronomie, the European Southern 
Observatory, and the Onsala Space Observatory.} observations of [CII] line emission in BRI~1335-0417. 
We adopt the following $\Lambda$-dominated cosmological parameters:
 $H_0 = 71$ km s$^{-1}$ Mpc$^{-1}$, $\Omega_\Lambda = 0.73$, $\Omega_m = 0.27$
(Spergel et al.\ 2007).

\section{Observations and Data Analysis}

Observations of [CII] line emission in BRI~1335-0417 were made with the 12m APEX telescope in November and December of 2009. The data were obtained over the course of five nights, as listed in Table~1, for a total of 20.3~hours of observing time. The weather conditions were generally dry, with a median precipitable water vapour, PWV$\sim$0.5~mm. 
At a redshift of $z = 4.4074$, determined from the \cofive emission line observed in BRI~1335-0417 by Guilloteau et al.\ (1997), the 1900.539~GHz [CII] line is redshifted to 351.47~GHz, accessible with the APEX-2 receiver. For this frequency,  we assume a gain of 41~Jy/K in converting $T_a^*$ to flux density (G{\"u}sten et al.\ 2006) and the absolute calibration uncertainties are expected to be 12\%.
 The Fast Fourier Transform Spectrometer was set up to sample $\sim$1000~km~s$^{-1}$ of bandwidth (two overlapping 1~GHz units), sufficient to cover the [CII] velocity width expected from the \cofive line ($\sim$420~km~s$^{-1}$; Guilloteau et al. 1997), assuming that the two species arise within a common region. The beamsize of these APEX-2 observations is 17.8$''$, much larger than the physical extent of the CO line emission (Riechers et al.\ 2008). 

\begin{table}[ht]
\centering
\caption{[CII] observations of BRI~1335-0417.}
\begin{tabular}{l c}
\hline \hline
Frequency:& 351.47 GHz  \\
Dates:  & 2009 Nov 29,  
                 2009 Nov 30, \\
               & 2009 Dec 1, 
                 2009 Dec 5, \\
               & 2009 Dec 6 \\
Pointing Center (J2000): & 13$^h$38$^m$03$^s$.38, -4$^o$32$^m$35$^s$.3  \\
rms (75~km~s$^{-1}$):   & 16.4 mJy  \\
\hline
\end{tabular}
\label{table1}
\end{table}

Data were analyzed using the GILDAS CLASS software package. Before creating the final, average spectrum, each scan was inspected for large frequency scale spectral baseline irregularities. After removing scans that clearly contained evidence for large scale instabilities, the total on-source integration time of scans contributing to the final spectrum is 214.1~minutes. Much of the large overhead cost associated with these observations is due to the use of the wobbling subreflector, essential for obtaining stable spectral baselines. 
Before adding the spectra, a constant, zeroth order baseline was subtracted from the spectrum of each scan, determined by fitting across the off line channels. Such a conservative baseline fitting approach was adopted so as to avoid introducing spurious line emission into the final spectrum. 
 Although there may be some uncertainty in the baseline subtraction, there is sufficient bandwidth for us to remove the continuum emission.

\section{Results}

   \begin{figure}[h]
   \centering
  \includegraphics[scale=0.4]{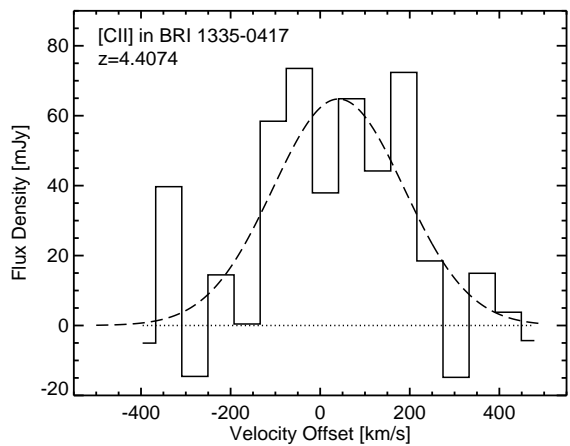}
   \caption{APEX spectrum of [CII] line emission redshifted to 351.47~GHz where the velocity scale is plotted relative to the \cofive redshift 
 $z = 4.4074$. The spectrum is plotted at a resolution of 58~km~$s^{-1}$ and the \textit{long dashed} line shows the fitted Gaussian with parameters listed in Table~2.}
              \label{Fig1}%
    \end{figure}

Figure~1 shows the APEX spectrum of [CII] line emission in BRI~1335-0417 where the continuum emission (expected to be $\sim$14~mJy; McMahon et al.\ 1999) is absent due to the baseline subtraction.
The rms in the final spectrum is $\sim$0.4~mK (16.4~mJy) per 75~km~s$^{-1}$ channel, calculated from the off-line channels away from the edge of the band. This is consistent with the predicted rms of 0.36~mK, estimated using 
  the total on-source observing time and a PWV of 0.5~mm.
The [CII] line emission is observed in the central channels of the spectrum, with a total integrated intensity of 26.6$\pm$4.3~Jy~km~s$^{-1}$. 
Following the definition for line luminosity given by equation (1) in Solomon, Downes \& Radford (1992), we calculate a [CII] line luminosity of 
$L_{[CII]} = (16.4 \pm 2.6)\times 10^9$~$L_{\odot}$ for BRI~1335-0417. Assuming that a single Gaussian profile is a good approximation to the [CII] line, we determine the best-fit parameters presented in Table~2. The line width and central frequency are consistent with that of the previously detected  \cofive line, measured at $z=4.4074$ with a linewidth of 420$\pm$60~km~s$^{-1}$. 

\begin{table}[ht]
\centering
\caption{[CII] line parameters for BRI~1335-0417.}
\begin{tabular}{ l  c }
\hline \hline
Parameter   & Value  \\
\hline
[CII] peak: & 65$\pm$19~mJy \\
$\Delta V_{\rm FWHM}$:  &  340$\pm$140~km~s$^{-1}$ \\
$v_0$$^a$:  &  41$\pm$61~km~s$^{-1}$ \\
$I_{[CII]}$:  &  26.6$\pm$4.3~Jy~km~s$^{-1}$ \\
$L_{[CII]}$:  & (16.4$\pm$2.6)$\times 10^{9}$~$L_{\odot}$ \\
\hline
\end{tabular}

$^a$Velocity with respect to $z=4.4074$, determined from the \cofive 
line Guilloteau et al.\ (1997).
\label{table2}
\end{table}

It has been proposed that the [CII] line luminosity can be used to estimate the obscured star-formation rate (SFR) in ultra- and hyper-luminous 
infrared galaxies such as BRI~1335-0417. Maiolino et al.\ (2005) use the calibration between FIR luminosity and SFR given by Kennicutt (1998)  to derive a relationship between $L_{[CII]}$ and SFR for systems with $L_{FIR} > 10^{12}$~$L_{\odot}$. We note that Boselli et al.\ (2002) find a lower conversion factor to infer SFR from $L_{[CII]}$. This is because the calibration of the relationship is based on the FIR luminosity, and it is observed that higher luminosity systems exhibit a decreasing [CII]-to-FIR luminosity ratio (Malhotra et al.\ 1997; Maiolino et al.\ 2009). Given the large apparent FIR luminosity in BRI~1335-0417, we adopt equation (1) from Maiolino et al.\ (2005) to estimate a SFR$\sim 10,700$~$M_{\odot}$~yr$^{-1}$ (but see the discussion in \S 4.1). 

\section{Discussion}

\subsection{[CII] line emission, FIR luminosity, and star-formation activity}

Although the  star-formation rate inferred from the [CII] line luminosity is extreme, the applicability of the relationship 
used here depends on whether BRI~1335-0417 has a similar [CII]-to-FIR luminosity ratio as the other ultra- and hyper-luminous infrared
galaxies from which this relationship is calibrated. The FIR luminosity determined by Benford et al.\ (1999) is $L_{FIR} = 3.1\times10^{13}~L_{\odot}$ and implies an obscured star-formation rate of 4650~$M_{\odot}~yr^{-1}$ following Kennicutt (1998), if all of the FIR luminosity 
 is powered by star-formation rather than AGN activity. Indeed, the AGN must be in a low state of activity as the brightness temperature 
measured from 1.4~GHz VLBI imaging is only $\sim$3.5$\times$10$^4$~K, and extended over scales of 1--2~kpc  (Momjian et al.\ 2007), reminiscent of synchrotron emission arising from star-forming galaxies and not a central AGN.
 From the [CII] line luminosity measured in BRI~1335-0417, we calculate, 
$L_{[CII]}/L_{FIR} = (5.3\pm0.8)\times 10^{-4}$, which is $\sim$3$\times$ higher than expected if one adopts the luminosity relationship determined
 for the other similarly luminous high-redshift objects detected in [CII] line emission (Maiolino et al.\ 2009). This would suggest that the
 star-formation rate inferred from the [CII] line luminosity following the Maiolino et al.\ (2005) relation is likely an overestimate of the 
true rate by a similar factor ($\sim$3) in the case of BRI~1335-0417.

Given the large scatter in the [CII]-to-FIR luminosity relationship observed in luminous and ultra-luminous infrared galaxies, 
 we do not consider the observed [CII] luminosity in BRI~1335-0417 to be unusual. This ratio has been shown to depend on dust temperature rather than FIR luminosity (Malhotra et al.\ 2001), which we have not considered for the high-redshift sample. 
It is possible that AGN heating of the dust plays a significant role here, however both J1148+5251 and BRI~0952-0115 are believed to contain AGN, and they contribute to the calibration of the high luminosity end of the [CII]-to-FIR luminosity relationship. 
The radio-to-FIR spectral energy distribution of BRI~1335-0417 is very similar to that of  the nuclear starburst galaxy M82 (Carilli et al.\ 1999), further suggestive of a starburst origin for the FIR luminosity in this object. 
  Sub-arcsecond resolution imaging of both the dust continuum and 
 [CII] line emission would enable us to measure the spatial distribution of these two components of the interstellar
 medium, and may help determine if the AGN plays a significant role in heating the dust. 

\subsection{Physical conditions of the gas}

Comparing the line luminosities in different molecular and atomic species can provide useful constraints on the cooling budget and conditions of the star-forming gas within BRI~1335-0417.
Walter et al.\ (\textit{in prep.}) obtain sensitive limits to the luminosity in the CI($^3P_2$-$^3P_1$) and CI($^3P_1$-$^3P_0$)
lines emitting from BRI~1335-0417. The upper-limit to the luminosity in the CI($^3P_1$-$^3P_0$) transition indicates that this line has less than 2\% of the luminosity in the [CII] line detected here. This is consistent with observations of the $z=6.4$ quasar J1148+5251, for which the CI($^3P_2$-$^3P_1$) line contributes $\sim$2\% of the ISM gas cooling compared to the [CII] emission line (Maiolino et al.\ 2005; Walter et al.\ 2009; Riechers et al.\ 2009). 

We can also compare the [CII] line luminosity to that of the \cotwo line emission used to infer the total mass of the cold molecular gas
 (Carilli et al.\ 1999, 2002;  Riechers et al.\ 2008). The intensity ratio between the high and low order CO lines suggests that the CO excitation is similar to that of BRI~1202-0725, whose gas has a kinetic temperature, $T_{kin} \sim$60~K and 
 a density, 10$^{4.1}$cm$^{-3}$ (Riechers et al.\ 2006, 2008). High resolution imaging of the \cotwo line emission reveals at least three components, indicative of an ongoing merger between multiple gas-rich systems (Riechers et al.\ 2008). If we assume a constant brightness temperature ratio between the \textit{J}=2-1 and \textit{J}=1-0 CO line transitions, consistent with that observed in other high-redshift quasar host galaxies (e.g. Riechers et al.\ 2006), then we can calculate the line lumiosity ratio of the \co and the [CII] lines for comparison with other galaxies. We calculate a line luminosity ratio, $L_{[CII]}/ L_{CO(J=1-0)} = 2900 \pm 480$. 

The [CII]-to-CO line luminosity ratio in BRI~1335-0417 is similar to that of the $z=4.7$ quasar host galaxy, BR1202-0725, however the [CII]-to-FIR 
luminosity ratio is larger in BRI~1335-0417, as discussed previously. From these luminosity ratios, models for gas in PDR regions  (e.g. Kaufman et al.\ 1999) can be used to constrain the gas density ($n$) and FUV flux ($G_0$, expressed in units of 1.6$\times$10$^{-3}$~ergs~cm$^{-2}$~s$^{-1}$).  The observations of BRI~1335-0417 are consistent with dense gas ($n \sim$10$^5$~cm$^{-3}$) illuminated by a FUV flux, $G_0 \sim 10^{3-4}$. These luminosity ratios and inferred physical conditions are similar to those found in BR1202-0725 as well as local ULIRGs (e.g. Hailey-Dunsheath et al.\ 2010). It would therefore seem that BRI~1335-0417 is indeed an extreme version of a local starburst ULIRG.

\section{Summary}

We have used the APEX telescope to observe the 157.74$\mu$m atomic cooling line of [CII] in the $z = 4.4074$ quasar host galaxy BRI~1335-0417. The line is detected with an apparent luminosity of $L_{[CII]} = (16.4 \pm 2.6)\times 10^9$~$L_{\odot}$,  making it the most intrinsically luminous [CII] line emitter so far at high-redshift. The high luminosity results in a superior [CII]-to-FIR luminosity ratio, $(5.3\pm 0.8)\times$10$^{-4}$, when compared to other high-redshift galaxies of comparable luminosities. This is consistent with the picture that  BRI~1335-0417 is an extremely luminous starburst galaxy in the process of undergoing a `wet' merger.

With the future capabilities of ALMA, observations of redshifted FIR emission lines will be possible for less luminous, high-redshift star-forming galaxies. For example, detecting the [CII] line emission from a $z=4.5$ object with a line strength $\sim$100$\times$ weaker than that of BRI~1335-0417\footnote{A 0.7~mJy peak [CII] line would correspond to,   SFR$\sim 2$~$M_{\odot}$~yr$^{-1}$, following the Bosselli et al. (2002) calibration.} would require only $\sim$4 hours of observing with the array of 12m ALMA antennas. Surveys of [CII] line emission with ALMA would therefore allow us to probe the dynamics and obscured star-formation in objects whose star-formation rates were comparable to that of the Milky Way when the Universe was only 1.4~Gyr old. 

\begin{acknowledgements}
We thank the staff of the APEX  telescope for help with the observations and data analysis. In particular we thank Thomas Stanke, Andreas Lundgren, 
Rodrigo Parra, Francisco Montenegro, Giorgio Siringo, and Claudio Agurto. DR acknowledges support from from NASA
through Hubble Fellowship grant HST-HF-51235.01 awarded by STScI, operated by AURA for NASA, under contract NAS 5-26555. We thank the referee for useful comments and suggestions on the manuscript.
\end{acknowledgements}


\begin{thebibliography}{} 

\bibitem{} Benford, D.~J., Cox, P., Omont, A., Phillips, T.~G., \& McMahon, R.~G.\ 1999, \apjl, 518, L65 
\bibitem{} Boselli, A., Gavazzi, G., Lequeux, J., \& Pierini, D. 2002, A\&A, 385, 454
\bibitem{} Carilli, C.~L., Menten, K.~M., \& Yun, M.~S.\ 1999, \apjl, 521, L25 
\bibitem{} Carilli, C.~L., et al.\ 2002, \aj, 123, 1838 
\bibitem{} Crawford, M.~K., Genzel, R., Townes, C.~H., \& Watson, D.~M.\ 1985, \apj, 291, 755 
\bibitem{} Guilloteau, S., Omont, A., McMahon, R.~G., Cox, P., \& Petitjean, P.\ 1997, \aap, 328, L1 
\bibitem{} G{\"u}sten, R., Nyman, L.~{\AA}., Schilke, P., Menten, K., Cesarsky, C., \& Booth, R.\ 2006, \aap, 454, L13 
\bibitem{} Iono, D., et al.\ 2006, \apjl, 645, L97 
\bibitem{} Ivison, R.~J., et al.\ 2010, arXiv:1005.1071 
\bibitem{} Kaufman, M.~J., Wolfire, M.~G., Hollenbach, D.~J., \& Luhman, M.~L.\ 1999, \apj, 527, 795 
\bibitem{} Kennicutt, R. C. 1998, ARA\&A, 36, 189
\bibitem{} Maiolino, R., et al.\ 2005, \aap, 440, L51 
\bibitem{} Maiolino, R., Caselli, P., Nagao, T., Walmsley, M., De Breuck, C., \& Meneghetti, M.\ 2009, \aap, 500, L1 
\bibitem{} Malhotra, S., et al.\ 1997, \apjl, 491, L27 
\bibitem{} Malhotra, S., Kaufman, M. J., Hollenbach, D., et al. 2001, ApJ, 561, 766
\bibitem{} McMahon, R.~G.,  Priddey, R.~S., Omont, A., Snellen, I., \& Withington, S.\ 1999, \mnras, 309, L1 
\bibitem{} Momjian, E., Carilli,  C.~L., Riechers, D.~A., \& Walter, F.\ 2007, \aj, 134, 694 
\bibitem{} Omont, A., McMahon, R.~G., Cox, P., Kreysa, E., Bergeron, J., Pajot, F., \& Storrie-Lombardi, L.~J.\ 1996, \aap, 315, 1 
\bibitem{} Riechers, D. A., et al. 2006, ApJ, 650, 604 
\bibitem{} Riechers, D.~A., Walter, F., Carilli, C.~L., Bertoldi, F., \& Momjian, E.\ 2008, \apjl, 686, L9 
\bibitem{} Riechers, D.~A., et  al.\ 2009, \apj, 703, 1338 
\bibitem{} Solomon, P.~M., Downes, D., \& Radford, S.~J.~E.\ 1992, \apjl, 398, L29 
\bibitem{} Spergel, D. N. et al. 2007, ApJS, 170, 377 
\bibitem{} Stacey, G.~J., Geis, N., Genzel, R., Lugten, J.~B., Poglitsch, A., Sternberg, A., \& Townes, C.~H.\ 1991, \apj, 373, 423 
\bibitem{} Storrie-Lombardi, L.~J., McMahon, R.~G., Irwin, M.~J., \& Hazard, C.\ 1996, \apj, 468, 121 
\bibitem{} Walter, F., Riechers, D., Cox, P., Neri, R., Carilli, C., Bertoldi, F., Weiss, A., \& Maiolino, R.\ 2009, \nat, 457, 699 
\bibitem{} Wright, E.~L., et al.\ 1991, \apj, 381, 200 



\end{thebibliography}
\end{document}